\documentclass[twocolumn,amsmath,amssymb,prl]{revtex4}
\usepackage{graphicx}
\usepackage{dcolumn}
\usepackage{bm}
\begin{document}
\title{Nonlinear coupling of nano mechanical resonators to Josephson 
quantum circuits}
\author{Xingxiang Zhou} 
\email{xizhou@yahoo.com} 
\author{Ari Mizel}
\email{ari@phys.psu.edu}
\affiliation{ Department of Physics, The Pennsylvania State 
University, University Park, PA 16802
}
\date{\today}

\begin{abstract}
We propose a technique to couple the position operator of a nano
mechanical resonator to a SQUID device by modulating its magnetic flux
bias. By tuning the magnetic field properly, either linear or
quadratic couplings can be realized, with a discretely adjustable
coupling strength. This provides a way to realize coherent nonlinear
effects in a nano mechanical resonator by coupling it to a Josephson
quantum circuit. As an example, we show how squeezing of the nano
mechanical resonator state can be realized with this technique. We
also propose a simple method to measure the uncertainty in the
position of the nano mechanical resonator without quantum state
tomography.

\end{abstract}

\pacs{85.85.+j, 85.25.Cp, 03.67.-a}

\maketitle

{\it Introduction.} Historically, mechanical systems have not been the
favorite proving ground of quantum mechanics because of their macroscopic
nature. This situation has changed in recent years thanks to the
impressive advance of nano fabrication technologies. It is now
possible to make nano mechanical resonators with frequencies of
gigahertz \cite{ref:Huang03} and quality factors approaching $10^5$ at
milli Kelvin temperatures \cite{ref:NEMS}. This opens the 
possibility of studying coherent quantum behavior in mesoscopic mechanical 
systems \cite{ref:Armour02,ref:Tian04}. 
One can also exploit 
the quantum properties of 
mechanical degrees of freedom for applications in areas such as weak
force detection \cite{ref:Bocko96}, precision measurement
\cite{ref:Munro02},
and quantum information processing \cite{ref:Cleland04}. 
Recently, evidence for quantized displacement in a nano mechanical
resonator has been observed \cite{ref:Gaidarzhy05}.

All micro and nano mechanical devices require some means of
transduction. As discussed in \cite{ref:Armour02}, 
an excellent way to engineer and detect the quantum modes of a nano
mechanical resonator is to couple the resonator to
Josephson device based solid state circuitry \cite{ref:Makhlin01} 
on which quantum coherent control has been demonstrated. 
A straightforward coupling scheme for this purpose 
is to voltage bias the resonator 
and use the position dependent electrostatic interaction between the
nano resonator and a charge island. This is discussed in
\cite{ref:Armour02} and used in all previous studies (see
\cite{ref:Armour02_cited} for an incomplete list). Though conceptually
simple and practically feasible, this scheme has the limitation that
the dominant term in the induced coupling is always linear in the
resonator position.  Consequently, it can only be used to realize
linear effects in the nano resonator. As is well known in
quantum optics \cite{ref:QO}, nonlinear effects 
are indispensable for thorough study and control of the dynamics of
harmonic oscillators. They are required to produce essential processes
such as squeezing and parametric amplification. They are also known to
be necessary for universal quantum information processing
on continuous variables \cite{ref:Braunstein05}. 
Thus, it is highly desirable to develop new coupling schemes which
can introduce nonlinear effects into the nano resonator system.

In this work, we propose an attractive alternative method to couple
the nano mechanical resonator to a Josephson quantum circuit
by modulating the flux bias of a SQUID device.
Our scheme makes it possible to realize both linear and nonlinear processes on
the nano mechanical resonator. We show how we can generate squeezing
of the nano mechanical resonator using this method and propose a
simple way to measure the reduction in the uncertainty in the
resonator's position.

\begin{figure}[h]
    \centering
    \includegraphics[width=2.6in, height=1.2in]{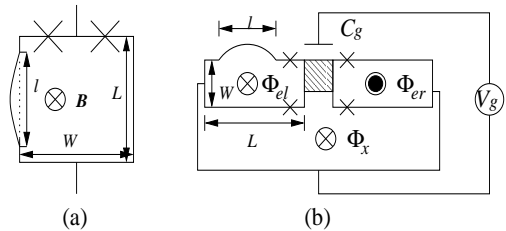}
    \caption{(a) A nano mechanical resonator in one arm of a SQUID device. 
      The SQUID loop is threaded by a magnetic field. Its flux bias
      will then depend on the position of the nano mechanical
      resonator. (b) A more complicated design in which a nano
      mechanical resonator is coupled to a charge qubit. The shaded
      box is the superconducting island of the charge qubit.  }

    \label{fig:NEMS_SQUID}
\end{figure}

{\it The device and its working principle.} 
Our scheme is illustrated in Fig. \ref{fig:NEMS_SQUID} (a).  Here, the
nano mechanical resonator is in one arm of a SQUID device. The SQUID
loop is biased with a perpendicular magnetic field $B$, such that the
flux threading the SQUID is the product of $B$ and the area of the
SQUID loop.

We see that the area of the SQUID loop is dependent on the position of
the nano mechanical resonator. In Fig. \ref{fig:NEMS_SQUID} (a), we
denote the width and length of the SQUID loop $W$ and $L$.  The length
and small displacement of the nano mechanical resonator are $l$ and
$X$, defined such that the area of the SQUID loop is $WL+lX$.
The total flux bias of the SQUID loop is $\Phi_e^0 + BlX$, where
$\Phi_e^0=BWL$ 
is the flux bias corresponding to the equilibrium position of the nano
mechanical resonator.
With the phase drops of the two junctions being $\phi_1$ and $\phi_2$,
the Josephson energy of the SQUID is $E_J=-E_J^0\cos\phi_1
-E_J^0\cos\phi_2 =-2E_J^0\cos(\phi_1-\phi_2)/2\cos(\phi_1+\phi_2)$,
where $E_J^0=I_c\Phi_0/2\pi$ is the Josephson energy of the
(identical) junctions, $I_c$ the junction critical current and
$\Phi_0$ the flux quantum. Since the superconducting order parameter
is singly valued, we must have $\phi_1-\phi_2 + 2\pi (\Phi_e^0 +
BlX)/\Phi_0 = 2\pi p$ for some integer $p$ leading to
\cite{ref:E_J_note}
\begin{equation}
E_J
=-2E_J^0\cos(\frac{\pi\Phi_e^0}{\Phi_0}+\frac{\pi Bl}{\Phi_0}X)\cos\phi, 
\label{eq:E_J}
\end{equation}
where $\phi=(\phi_1+\phi_2)/2$ is the average phase across the 
junctions.

It is clear from Eq. (\ref{eq:E_J}) that in general the Josephson
energy of the SQUID is a nonlinear function of $X$, the position of
the nano mechanical resonator.
Treating $X$ as a small perturbation, we find that there are two flux
bias points of particular interest: $\Phi_e^0=(n+1/2)\Phi_0$ and
$\Phi_e^0=n\Phi_0$, where $n$ is an integer.  When
$\Phi_e^0=(n+1/2)\Phi_0$, to lowest order in $X$ the Josephson energy
of the SQUID is $E_J=(-1)^n2E_J^0(\pi (n+1/2)l/WL)X\cos\phi$. In this
case, $E_J$ has a linear dependence on $X$ and thus linear coupling
between the nano resonator and SQUID is realized. On the other hand,
when the SQUID loop is biased at $\Phi_e^0=n\Phi_0$, to lowest order
in $X$ the Josephson energy is
\begin{equation}
E_J=-(-1)^n2E_J^0\{1- (\pi nl/WL)^2X^2/2\}\cos\phi. 
\label{eq:nonlinear}
\end{equation}
Here, the Josephson energy of the SQUID has a quadratic 
dependence on the position of the nano mechanical resonator. 

From the perspective of micro and nano mechanical engineering, our
coupling scheme can be considered a magnetic transduction method.
Compared to the electrostatic transduction scheme \cite{ref:Armour02},
the distinctive advantage of our scheme is that the coupling can be
either linear or quadratic, depending on the flux bias point of the
SQUID. Also, notice that the coupling strength can be adjusted
(discretely), since $E_J$ depends on $n$ as in Eq.
(\ref{eq:nonlinear}). It is then possible in principle to operate in
both linear and nonlinear, as well as both weak and strong, coupling
regimes. In practice, the SQUID will be part of a Josephson quantum
circuit (for instance a charge or flux qubit), and the modulation
scheme above then couples the nano mechanical oscillator to the
Josephson quantum circuit.

Note the nano mechanical resonator must be superconducting for our
scheme to work. This can be realized by using a metalized resonator
\cite{ref:NEMS,ref:Schwab02} as long as the magnetic field does not
exceed the critical field strength of the superconductor.

{\it Squeezing of the nano mechanical resonator.}  An important
nonlinear effect on a harmonic oscillator is squeezing. It is the
suppression of the uncertainty in the position (or momentum) of the
oscillator at the price of increased uncertainty in the conjugate
variable. For nano mechanical resonators, reduction in the position
uncertainty is important in applications such as weak force detection
\cite{ref:Bocko96}.  Also, some coherent nonlinear process such as
squeezing is essential for universal quantum information processing on
continuous variables \cite{ref:Braunstein05}.  Our nonlinear coupling
scheme makes such coherent nonlinear processes inaccessible by
previous methods \cite{ref:Armour02,ref:Armour02_cited} possible. This
provides a way of introducing nonlinear-effect-induced operators, for
instance, as gates within a complicated quantum circuit. Previously,
generation of certain squeezed states of nano mechanical resonators
has been studied by a few authors \cite{ref:Rabl04,ref:Ruskov05}.
However, since no coherent nonlinear transduction scheme on the nano
mechanical resonator was known, these schemes had to function
incoherently. They used dissipation and measurement to generate the
needed nonlinearity and were incapable of effecting unitary
evolutions.


In the following we show how we can realize squeezing on a nano
mechanical resonator using our scheme to couple it to a charge qubit
\cite{ref:Shnirman97}.
First, we notice that the simple prototype circuit in Fig.
\ref{fig:NEMS_SQUID} (a), though convenient in illustrating the
essence of our scheme, has a few disadvantages. In particular,
according to Eq. (\ref{eq:nonlinear}), when a nonlinear coupling is
realized the Josephson self-energy is also the largest ($2E_J^0$),
therefore we do not have completely independent control over the
coupling strength and Josephson self-energy. To overcome this and
other disadvantages, we consider the design shown in Fig.
\ref{fig:NEMS_SQUID} (b). Here, we have two identical SQUIDs ({\it
  l}eft and {\it r}ight) biased at equal but opposite fluxes
$\Phi_{el} =-\Phi_{er} =BWL$ (this can be realized by twisting the
arms of one of the SQUIDs, for instance). The nano mechanical
resonator is in the arm of one of the SQUIDs and coupled to its phase.
The big loop is biased with an external flux $\Phi_x$.
This structure bears some similarity to the scalable 
charge qubit scheme in \cite{ref:You02}, and the advantage is that both 
the flux biases of the individual SQUIDs and the big loop can be tuned. 

In order to realize nonlinear coupling between the nano mechanical
resonator and the charge qubit, we bias the SQUIDs at $\Phi_{el}
=-\Phi_{er} =n\Phi_0$.
Using the same argument that leads to Eq. (\ref{eq:nonlinear}), we can 
derive the Josephson energy of the circuit in Fig. \ref{fig:NEMS_SQUID} (b).
To lowest order in the resonator position $X$, 
$E_J=-(-1)^n 4E_J^0\cos(\pi\Phi_{x}/\Phi_0)\cos\phi +(-1)^n E_J^0 (\pi nl/WL)^2
X^2\cos\phi_l$, where $\phi_l$ and $\phi_r$ are the phases of the left and 
right SQUIDs (the averages of the phases of the two junctions in the SQUIDs), 
and $\phi= (\phi_l +\phi_r)/2$ is the average phase of the SQUIDs conjugate 
to the charge number on the island. If we bias the big loop at 
$\Phi_x= (2m+1/2)\Phi_0$, $m$ an integer, 
only the coupling term survives, so $E_J=(-1)^n E_J^0 (\pi nl/WL)^2
X^2\cos\phi_l$.  

The charge island possesses an adjustable gate voltage $V_g$ which is
applied through the gate capacitance $C_g$. When it is biased close to
$n_g=C_gV_g/2e=1/2$, the states with $0$ and $1$ excess Cooper pairs
comprise the low energy Hilbert space of the qubit. Considering these
charge states the spin up and spin down states \cite{ref:Armour02} in
an effective two state system, we can use the Pauli matrices to
describe operators acting on the system, and its uncoupled Hamiltonian
takes the form $E_z\sigma_z/2$, where $E_z =(2n_g-1)(2e)^2/2C_t$
and $C_t$ is the total capacitance of the charge island.

As usual, the nano resonator is treated as a harmonic oscillator with
position operator $X=\delta X_0 (a+a^\dagger)$ \cite{ref:Armour02},
where $a$ is the annihilation operator, $\delta X_0
=\sqrt{\hbar/2M\omega_0}$ is the zero point fluctuation in the
resonator's position $X$, and $M$ and $\omega_0$ are the mass and
frequency of the resonator.  Its uncoupled Hamiltonian is $\omega_0
a^{\dagger}a$.

The system Hamiltonian is then $H =E_z\sigma_z/2 +\omega_0
a^{\dagger}a -(\lambda_n/2) (a+a^{\dagger})^2 \sigma_y$, where the
coupling strength $\lambda_n =-(-1)^n E_J^0(\pi nl\delta X_0/WL)^2$.
If we choose $E_z= 2\omega_0$ and shift to the rotating frame defined
by $E_z\sigma_z/2 +\omega_0 a^{\dagger}a$, we obtain the following
Hamiltonian in the rotating frame \cite{ref:QO}:
\begin{equation}
H_R= i(\lambda_n/2)(a^2\sigma^{+} -a^{\dagger 2}\sigma^{-}) 
+H_{\omega_0},
\label{eq:H_RWA}
\end{equation}
where $\sigma^{\pm} =(\sigma_x\pm i\sigma_y)/2$ and $H_{\omega_0}$ are
off resonance terms of magnitude $\lambda_n/2$ oscillating at
frequencies of $2\omega_0$ and higher. Since the realizable coupling
strength $\lambda_n$ is usually much smaller than the resonator
frequency $\omega_0$, we can adopt the rotating wave approximation to
drop $H_{\omega_0}$ \cite{ref:QO}.


In addition to the operating 
point discussed above, we also consider another set of bias conditions 
in which 
the SQUIDs are biased at 0 flux and 
$E_z$ is changed (by tuning the gate voltage $V_g$) to 
$E_z' =2\omega_0+ \delta E_z$. 
We also bias the big loop slightly away from $(2m+1/2)\Phi_0$ using a small 
ac field, $\Phi_x= (2m+1/2)\Phi_0 +\delta \Phi_x\cos 2\omega_0 t$ where 
$\pi \delta\Phi_x/\Phi_0 \ll 1$. 
In this case the charge qubit is decoupled 
from the resonator. To lowest order in $\pi \delta\Phi_x/\Phi_0$, 
the system Hamiltonian in the same rotating frame is $H =\delta 
E_z\sigma_z/2 +E_x\sigma_x/2 +H'_{\omega_0}$, where $E_x 
= 4E_J^0(\pi \delta\Phi_x/\Phi_0)$ and the rapidly oscillating
term $H'_{\omega_0}$ 
will have negligible effect if $E_x$ is small compared to $4\omega_0$ 
and the system evolves for an appropriate duration  \cite{ref:Wei63}. 
In the above Hamiltonian both $\delta E_z$ and $E_x$ of the qubit can
be adjusted \cite{ref:Armour02}; therefore we can perform arbitrary
rotations on the state of the charge qubit. In particular, if we
choose large values for $\delta E_z$ and $E_x$, 
we can realize pulsed operations on the charge qubit and flip its
state quickly.

We now consider a spin echo like process in which we let the system
evolve under the Hamiltonian (\ref{eq:H_RWA}) for short periods of
time of duration $\Delta t$. In between each such time interval we
apply a quick $\pi$ pulse $\sigma_x$ to the charge qubit to flip its
state, so that the evolution for two periods is governed by
$\exp\{-iH_{R}\Delta t \}\sigma_x \exp\{-iH_{R}\Delta t\}\sigma_x$.
Since $\sigma_x \exp\{(\lambda_n \Delta t/4) [(a^2-a^{\dagger
  2})\sigma_x + i(a^2+a^{\dagger 2}) \sigma_y]\} \sigma_x$ $=
\exp\{(\lambda_n \Delta t/4) [(a^2-a^{\dagger 2})\sigma_x -
i(a^2+a^{\dagger 2}) \sigma_y]\}$, this evolution operator can be
simplified: $\exp\{-iH_{R}\Delta t \}\sigma_x \exp\{-iH_{R}\Delta
t\}\sigma_x \approx \exp\{(\lambda_n \Delta t/2) (a^2-a^{\dagger 2})
\sigma_x\}$.  If we initialize the charge qubit in the $\sigma_x =1$
state, and repeat this procedure $N$ times, the evolution operator on
the state of the resonator becomes
\begin{equation}
\label{squeezeop}
S(\kappa)=\exp\{\frac{\kappa}{2} (a^2 -a^{\dagger 2})\},
\end{equation}
where $\kappa =\lambda_n N\Delta t$. 
This is a squeezing operator
on the nano mechanical resonator with squeezing parameter 
$\kappa$. Under the squeezing operator, $a$ transforms to
$S^\dagger(\kappa)aS(\kappa)=a\cosh\kappa-a^\dagger\sinh(\kappa)$ and it
can be shown that the position uncertainty decreases exponentially 
\cite{ref:QO}: $\Delta X= \sqrt{\langle X^2\rangle 
-\langle X\rangle^2}= \Delta X(0)e^{-\kappa}$.




In the above, we used a multi-loop circuit which allows 
bias and control of the individual loops. Such structure and control 
are easy to realize and widely used in current Josephson quantum
circuit design and experiment \cite{ref:You02, ref:Mooij99, ref:Mooij03}. 
Also, high-precision spin echo control 
of superconducting qubits has been realized experimentally 
\cite{ref:ChargeEcho}. Sophisticated microwave pulse sequences can be 
applied and it was observed that the decoherence time of the superconducting
qubit increases due to the spin echo control. Therefore, our scheme is  
within the reach of current technologies. 

Ideally, one would like a large coupling strength in order to generate
appreciable squeezing in a short period of time.
For a magnetic field $B \approx 0.2$T, $l \approx 30\mu$m, $\delta
X_0 \approx 5 \times 10^{-13}$m, and a critical current of about $60
$nA, the coupling strength $\lambda_n$ is about $4$MHz.

\begin{figure}[h]
    \centering
    \includegraphics[width=3.2in, height=1.2in]{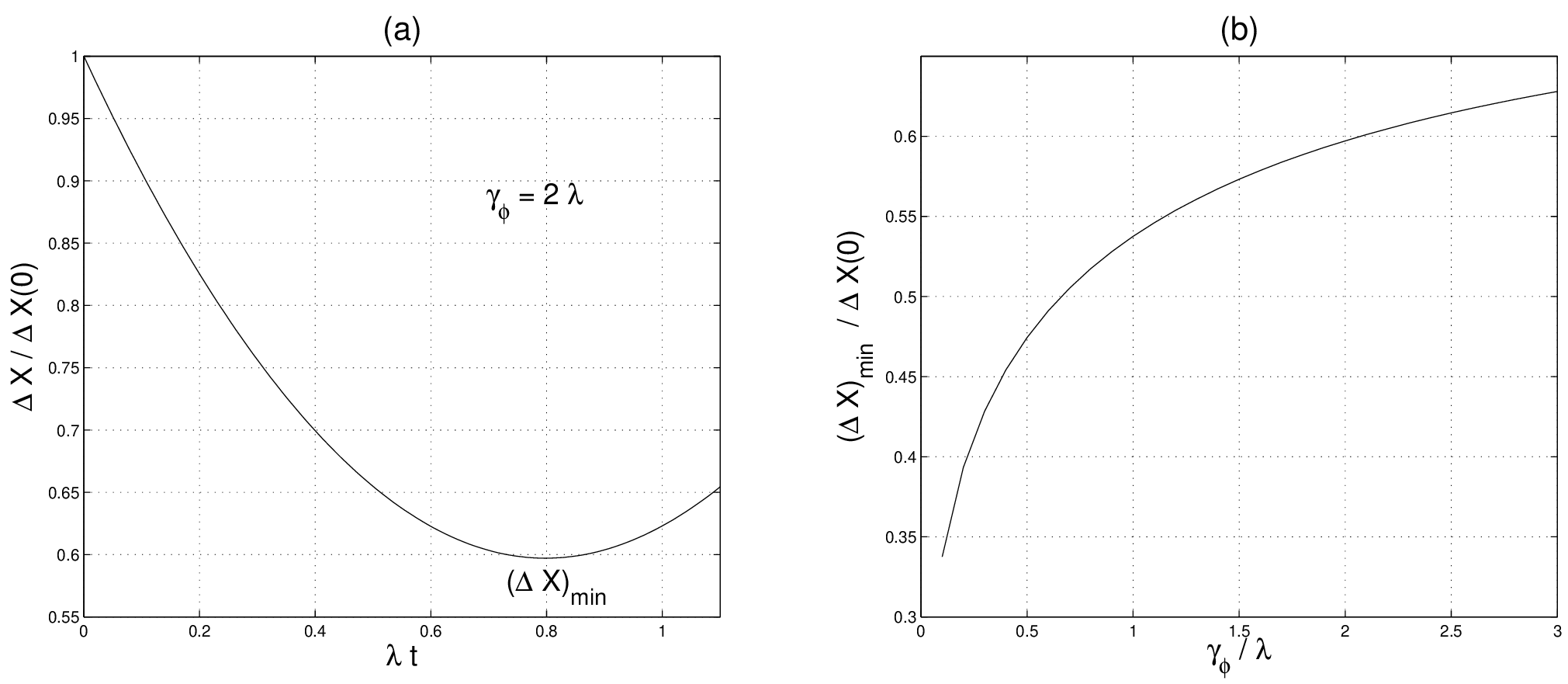}
    \caption{(a) Time dependence of the nano resonator position uncertainty
      $\Delta X$ (normalized to the initial uncertainty) under the
      effect of squeezing and decoherence. See text for experimental
      parameters and initial conditions.  (b) The maximum achievable
      squeezing (minimum $\Delta X$) as a function of the dephasing
      rate. All other parameters are the same as in (a).  }

    \label{fig:Delta_X}
\end{figure}

{\it Decoherence}.  Unlike previous methods
\cite{ref:Rabl04,ref:Ruskov05}, our scheme effects a coherent quantum
process. The influence of decoherence must be considered in detail.
For this purpose, we use the Master equation \cite{ref:QO}
\begin{equation}
d\rho/dt=-i[H_{sq}, \rho]+
\mathcal{L}(a,\gamma_n,N_n) +\mathcal{L}(\sigma^-,\gamma_q,N_q) +
\mathcal{L}(\sigma_z,\gamma_\phi,N_q), 
\end{equation}
where $\rho$ is the density matrix of the nano resonator - charge
qubit system, $H_{sq}=i\lambda(a^2-a^{\dagger 2})\sigma_x/2$ is the
effective squeezing Hamiltonian, $\gamma_n=\omega_0/Q$ is the decay
rate of the nano mechanical resonator determined by its quality factor
$Q$, $\gamma_q$ and $\gamma_\phi$ are the relaxation and dephasing
rate of the charge qubit, $N_n$ and $N_q$ are the mean values of the
bath quanta dependent on temperature \cite{ref:QO}, and the
Liouvillian operator $\mathcal{L}(A, \gamma, N)=
(1/2)\gamma(N+1)(2A\rho A^\dagger-A^\dagger A \rho -\rho A^\dagger A)+
(1/2)\gamma N(2A^\dagger\rho A-AA^\dagger\rho -\rho AA^\dagger)$.
Among the various decoherence sources, the 
dephasing of the charge qubit is dominant. For most nano mechanical
resonator frequencies \cite{ref:NEMS}, we only bias the charge qubit
slightly away from the charge degeneracy point ($|2n_g-1| \ll 0.1$)
and the dephasing time $T_2$ is greater than $100$ns
\cite{ref:Moon05}. Using the Master equation we can derive equations
for the expectation values of the dynamic variables of the system
which are then solved. In Fig.  \ref{fig:Delta_X} (a), we plot the
time dependence of the nano resonator position uncertainty $\Delta X$.
We have used the following conservative set of experimental
parameters: resonator frequency $\omega_0/2\pi=250$MHz, quality factor
$Q=10^4$, temperature $T=20$mK, squeezing parameter $\lambda=5$MHz,
with $\gamma_q$ and $\gamma_\phi$ chosen such that the relaxation time
$T_1=1\mu$s and dephasing time $T_2=100$ns \cite{ref:Moon05}.
Initially the nano resonator is assumed to be in a thermal equilibrium
state and the charge qubit is in the $\sigma_x=1$ state.

It is clear from Fig. \ref{fig:Delta_X} (a) that appreciable squeezing
can be generated even when the charge qubit's dephasing rate is severe
(twice the squeezing parameter), indicating that the squeezing is
robust against decoherence. As time progresses, the squeezing becomes
less effective and eventually $\Delta X$ starts to increase.  This is
easy to understand from the effective Hamiltonian
$i\lambda(a^2-a^{\dagger 2})\sigma_x$ leading to (\ref{squeezeop});
as the charge qubit dephases, $\langle \sigma_x\rangle$ decreases and
the squeezing effect weakens and eventually disappears. Not
surprisingly, the maximum squeezing achievable increases with decreasing
dephasing rate, as shown in Fig.  \ref{fig:Delta_X} (b).

{\it Measurement.}  Once we generate squeezing on the nano mechanical
resonator, a scheme is needed to measure the uncertainty in its
position to confirm the squeezing effect. One way to do this is
quantum tomography which measures the resonator's Wigner function.
As shown in \cite{ref:Lutterbach97}, the Wigner function can
be measured by displacing the state of the harmonic oscillator,
letting it interact with a two state system, and measuring the
polarization of the two state system. Though this method can determine
all the information about the oscillator, its direct application in
our solid state system is hindered by technical difficulties.  To
displace the resonator's state arbitrarily,
we will need couplings to both the position and momentum operators of
the resonator with continuously variable relative strengths, which is
not easy to realize in the nano resonator. Also, in order to calculate
$\Delta X$, we need the Wigner function over its entire parameter
space making it necessary to sweep through a large parameter space.
Large displacement will take a long time to effect and the decoherence
will affect the measurement result.

Here we propose a simplified method to measure the mean and variance
of the position operator of the nano resonator. It is based on the
measurement of the generating function $Tr(\rho e^{i\kappa X})$. We
notice that $\langle X \rangle$ and $\langle X^2\rangle$ can be
calculated from the generating function by
\begin{equation}
\langle X \rangle = Tr(\rho X) = -i\frac{d}{d\kappa}Tr(\rho 
e^{i\kappa X}) |_{\kappa=0}
\label{eq:X}
\end{equation}
and 
\begin{equation}
\langle X^2 \rangle = Tr(\rho X^2) = -\frac{d^2}{d\kappa^2}Tr(\rho 
e^{i\kappa X}) |_{\kappa=0}.
\label{eq:X_squared}
\end{equation}
Therefore, measuring the generating function in the vicinity of $\kappa =0$ 
allows us to determine the position uncertainty of the nano mechanical 
resonator $ \Delta X^2 $. 

The generating function can be measured
by first  preparing the charge qubit in the $\sigma_z=1$ state. Then turning on a strong 
coupling $\lambda X\sigma_x$ between the charge qubit and the resonator, for a time $t$,
will cause the charge qubit states corresponding to $\sigma_x=\pm
1$ to acquire a phase shift $\mp \lambda tX$. Next, the interaction is turned off and the gate
$\textsf{H}\exp\{i(\pi/4)(\vec{n}\cdot \vec{\sigma})\}$ is applied to the charge
qubit, where $\textsf{H}= (\sigma_x+\sigma_z)/\sqrt{2}$ is the
Hadamard gate, $\vec{n} =(0, -\cos{\eta}, \sin{\eta})$, $\eta$ an
angle chosen to be 0 or $\pi/2$. We then measure the polarization of
the charge qubit, $\langle \sigma_z \rangle = P_{\sigma_z=1}
-P_{\sigma_z=-1}$, which can be shown to equal 
\cite{ref:Lutterbach97}
\begin{equation}
Tr (\rho Re\{ e^{i\eta}e^{i\kappa X}\}),
\end{equation}
where $\kappa =2\lambda t$. Choosing $\eta=0$ and $\pi/2$ then 
yields 
the real and imaginary part of the generating function, which in turn 
allows us to calculate $ \Delta X^2 $ by Eqs. (\ref{eq:X}) 
and (\ref{eq:X_squared}). 

This method based on the measurement of the generating function has a few 
attractive advantages. It requires only strong linear coupling of the charge 
qubit to the position operator of the resonator, which as discussed before
can be easily realized 
in our scheme by biasing the SQUIDs at $(n+1/2)\Phi_0$ and the big loop 
at 0 flux.
We only need to measure in the vicinity of $\kappa =0$, 
therefore the measurement can be done quickly, implying less influence by
decoherence. Since the polarization of the charge qubit can be measured 
with high fidelity \cite{ref:Vion02}, our scheme is realistic given 
currently available technology.

{\it Conclusion.} We have proposed a scheme to couple a nano
mechanical resonator to Josephson quantum circuits by modulating the
magnetic bias of a SQUID. This allows us to realize coherent nonlinear
effects on the nano resonator, which are essential for but so far
missing in the study of nano mechanical resonators.  Though we focused
on the squeezing of a nano mechanical resonator by coupling it to a
charge qubit, our scheme can be easily tailored to other purposes and
adapted for coupling to other Josephson quantum circuits.  It can be
directly extended for quantum manipulation of multiple nano
resonators. Then, entanglement can be generated and confirmed using
the simple measurement scheme. This can provide a practically feasible
approach for unambiguous demonstration of quantum behavior in nano
mechanical resonators.

The authors acknowledge financial support from the Packard foundation.

\end{document}